\documentclass[aps,prb,two  column,superscriptaddress]{revtex4-1}
\usepackage{graphicx}
\usepackage{amssymb, amsmath}
 
\newcommand{\Irtwo}{$\mathrm{Sr}_2\mathrm{Ir}\mathrm{O}_4$}
\newcommand{\Latwo}{$\mathrm{La}_2\mathrm{Cu}\mathrm{O}_4$}

\newcommand{\Rhtwo}{$\mathrm{Sr}_2\mathrm{Rh}\mathrm{O}_4$}

\newcommand{\Rhdoped}{$\mathrm{Sr}_2\mathrm{Ir}_{1-x}\mathrm{Rh}_x\mathrm{O}_4$}

\begin{document}

\title{Charge partitioning and anomalous hole doping in Rh-doped \Irtwo}



\author{S. Chikara}
\affiliation{Advanced Photon Source, Argonne National Laboratory, Argonne, IL 60439, USA} \affiliation{ National High Magnetic Field Laboratory (NHMFL), Los Alamos National Laboratory, Los Alamos, New Mexico 87545, USA}

\author{G. Fabbris}
\affiliation{Advanced Photon Source, Argonne National Laboratory, Argonne, IL 60439, USA} \affiliation{ Department of Physics,
Washington University, St. Louis, Missouri 63130, USA} \affiliation{Department of Condensed Matter Physics and Materials Science, Brookhaven National Laboratory, Upton, New York 11973, USA} 

\author{J. Terzic}
\affiliation{Center for Advanced Materials and Department of Physics and Astronomy, University of Kentucky, Kentucky, Lexington, Kentucky 40506, USA}

\author{G. Cao}
\affiliation{Department of Physics, University of Colorado-Boulder, Boulder, Colorado 80309, USA}

\author{D. Khomskii}
\affiliation{Physikalisches Institut, Universit\"{a}t zu K\"{o}ln, Z\"{u}lpicher Strasse 77, 50937 K\"{o}ln, Germany }

\author{D. Haskel}
\thanks{Corresponding author; {haskel@aps.anl.gov}}
\affiliation{Advanced Photon Source, Argonne National Laboratory, Argonne, IL 60439, USA}


\begin{abstract}

The simultaneous presence of sizable spin-orbit interactions and electron correlations in iridium oxides has led to predictions of novel ground states including Dirac semimetals, Kitaev spin liquids, and  superconductivity. Electron and hole doping studies of spin-orbit assisted Mott insulator \Irtwo\ are being intensively pursued due to extensive parallels with the La$_2$CuO$_4$ parent compound of cuprate superconductors. In particular, the mechanism of charge doping associated with replacement of Ir with Rh ions remains controversial with profound consequences for the interpretation of electronic structure and transport data. Using x-ray absorption near edge structure (XANES) measurements at the Rh {\it L, K}- and Ir {\it L}- edges we observe anomalous evolution of charge partitioning between Rh and Ir with Rh doping. The partitioning of charge between Rh and Ir sites progresses in a way that holes are initially doped into the J$_{\rm eff}$=1/2 band at low $x$ only to be removed from it at higher $x$ values. This anomalous hole doping naturally explains the reentrant insulating phase in the phase diagram of \Rhdoped\ and ought to be considered when searching for superconductivity and other emergent phenomena in iridates doped with 4d elements.

\end{abstract}



\maketitle            


\noindent 

The strong spin orbit coupling (SOC) and the novel J$_{\rm eff}$=1/2 physics in iridates makes them a promising candidate for exotic physical phenomena.~\cite{Witczak-Krempa2014, Jackeli2009, Shitade2009, Pesin2010, Wang2011, Okada2013} The underlying signatures  of superconductivity in K-doped \Irtwo~\cite{Kim2014} has given impetus to the possibility of doping-driven superconductivity in \Irtwo. \Irtwo\ shows close parallels to the parent cuprate \Latwo\ with
a quasi-two-dimensional structure ~\cite{Crawford1994, Chikara2009} and Ir$^{4+}$ ions in a square lattice with one hole per Ir site. Furthermore, the low energy magnetic excitations are well described by the antiferromagnetic Heisenberg model~\cite{Kim2008, Kim2009, Kim2012, Fujiyama2012, Wang2011} and pseudo gap like states are observed in various doped systems.~\cite{Kim2014, Cao2016, DelaTorre2015} There have been theoretical proposals and experimental observations of underlying signatures of superconductivity in \Irtwo\ via both electron~\cite{Wang2011, Watanabe2013, Kim2014, Kim2015, Yan2015} and hole doping.~\cite{Meng2014, Kong2015} In this context, Rh doped \Rhdoped\ has attracted much attention as it displays pseudogap region in ARPES experiments~\cite{Cao2016} and hidden broken symmetry in second-harmonic generation experiments~\cite{Zhao2016} raising prospects for the possibility of superconductivity in \Irtwo\ doped with Rh or other 4d elements.

There have been conflicting reports on the oxidation state of Rh and nature of electronic doping in \Rhdoped. Some studies~\cite{Qi2012a} assume Rh enters the lattice as Rh$^{4+}$ (4d$^5$ configuration), i.e., isoelectronic to Ir$^{4+}$ (5d$^5$). In this case it introduces neither holes nor electrons and the transition to a metallic state at $x\sim 0.16$ is believed to be a consequence of reduced spin orbit interactions in the lighter 4d element. Alternatively, other studies report that Rh enters the lattice as Rh$^{3+}$ (4d$^6$ configuration) where an electron is transferred out of J$_{\rm{eff}}$=1/2 states thereby doping holes.~\cite{Clancy2014, Cao2016} Therefore, there is no clear  understanding on how Rh doping drives \Rhdoped\ from a Mott insulator to a metallic state. Since hole- and electron-doped cuprates present significant differences in how they lead to superconductivity,~\cite{Weber2010} it is important to know the exact nature of doping in this and other iridates in order to draw reliable parallels with the cuprates.

Several experiments have reported on the evolution of electronic structure 
in \Irtwo\ as a function of Rh doping.~\cite{Sohn2016, Cao2016, Ye2015, Brouet2015, Qi2012a, Clancy2014, Chikara2015} The substitution of Rh in \Irtwo\ rapidly suppresses antiferromagnetic order (T$_N$) for concentrations as low as 16 at.$\%$. The suppression of magnetic ordering is accompanied by a six orders of magnitude drop in the resistivity. Surprisingly, the system shows re-entrant insulating behavior for $x\geq 0.24$.~\cite{Qi2012a} The emergence of metallicity at low $x$ has been attributed, under assumption of isovalent substitution, to the collapse of the Mott gap driven by the smaller SOC of the lighter Rh.~\cite{Qi2012a}  However, a recent report shows no perturbation of the expectation value $\langle \bold{L}\cdot\bold{S}\rangle$ in the Ir $5d$ band with increasing $x$, appearing to invalidate this conclusion.~\cite{Chikara2015} The same report shows formation of impurity Rh bands overlapping with the lower Hubbard band (LHB) in \Irtwo\ as a likely cause for the insulator-metal transition~\cite{Chikara2015}. Angle-resolved photoemission (ARPES) measurements at low Rh doping show rigid band shift with no appreciable change in band dispersion, the chemical potential moving into the LHB consistent with hole doping of J$_{\rm eff}$=1/2 states~\cite{Cao2016}. Since valence state determines band filling, it is essential to understand the evolution of Rh and Ir oxidation state to address the mechanism of electronic doping as well as the reentrant insulating behavior.


We collected x-ray absorption near edge structure (XANES) data at Rh {\it K, L}- and Ir {\it L}-edges in \Rhdoped. Polycrystalline and single crystal samples of \Rhdoped\ ($0 \leq x \leq 0.70$) were prepared by a solid state reaction~\cite{Qi2012a} and characterized by powder x-ray diffraction to rule out impurity phases. The Rh content was validated with energy dispersive x-ray analysis (EDX) while thermogravimetric analysis (TGA) was used 
to test for changes in oxygen content in pure and representative Rh doped samples. We found that more than 5\% of oxygen depletion would result in decomposition of the samples. Given the unusual chemical stability of the iridates, the oxygen content prefers to be 4 in almost all cases we have studied.  The samples were ground  and sieved to $\approx 10\, \mu$m particle size for all measurements. The Ir $\rm{\it L}_{2, 3}$- (electric dipole transition $2p \rightarrow 5d$), Rh $\rm{\it L}_3$- ($2p \rightarrow 4d$) and Rh {\it K}-edge ($1s \rightarrow 5p$) XANES measurements were done at beamline 4-ID-D of the Advanced Photon Source at Argonne National Laboratory. The Ir $\rm{\it L}_{2, 3}$- and Rh {\it K}- edge measurements were performed in transmission geometry whereas Rh $\rm{\it L}_3$- edge measurements were performed by detecting the partial fluorescence yield of Rh $\rm{\it L}_{\alpha}$ emission with a 4-element energy resolving silicon drift diode detector as the x-ray energy was scanned through the Rh $\rm{\it L}_3$ absorption edge.

\begin{figure}
 \includegraphics[scale=0.28,angle= 0]{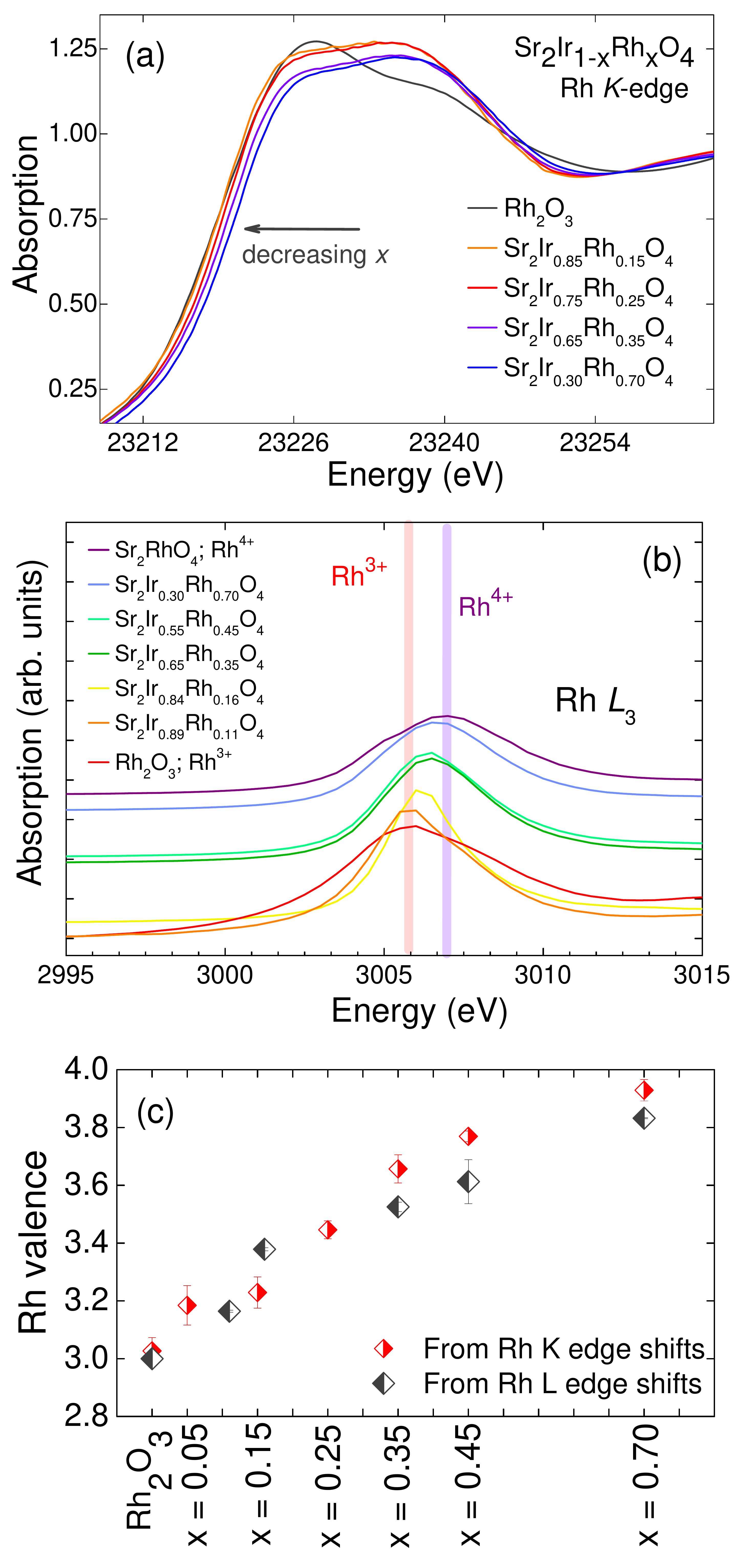}
 \caption{(a) Edge-jump normalized Rh {\it K}-edge XANES data for \Rhdoped\ and Rh$_2$O$_3$ reference, all measured at room temperature (b) XANES measurements at the Rh $\rm{\it L}_3$ edge for \Rhdoped\ with Rh$^{3+}$ and Rh$^{4+}$ reference samples. All plots are shifted in the vertical direction for clarity (c) Derived Rh oxidation state based on energy shifts of leading Rh {\it K} and {\it L} absorption edges.}
\label{RhKedge}
\end{figure}

Fig.~\ref{RhKedge}(a) shows normalized Rh {\it K}-edge XANES data for \Rhdoped\ samples and a Rh$^{3+}$ oxide standard. There is a systematic shift in the leading edge towards higher energy with increasing $x$ (shift also seen in the first derivative data). This shift is indicative of a change in Rh oxidation state.~\cite{Haskel2004, Huang2014, Croft1997} Normalized Rh $\rm{\it L}_3$ edge XANES data for various doping levels together with Rh$^{3+}$ and Rh$^{4+}$ oxide standards are shown in Fig.~\ref{RhKedge} (b). The curves are shifted vertically for clarity. The main absorption peaks of the 11\% and 70\% samples are close in energy to those of the Rh$^{3+}$ and Rh$^{4+}$ reference standards, respectively. Those of samples with intermediate $x$ values lie between the two standards, clearly marking the systematic change in Rh valence consistent with {\it K}-edge measurements. Fig.~\ref{RhKedge}(c) shows the oxidation state derived from the shift in leading edge of both Rh {\it K}- and {\it L}-edge data. The oxidation state is calculated using a linear interpolation based on the energy shift per Rh unit charge ($1.82 \pm 0.08$ eV for {\it K}-edge and $1.4 \pm 0.04$ eV for $\rm{\it L}_3$-edge ) obtained from Rh$_2$O$_3$(3+) and \Rhtwo(4+) standards. Clearly results from independent measurements at both edges are in good agreement. Contrary to earlier reports of Rh being in either Rh$^{4+}$ or Rh$^{3+}$ oxidation state, our results point towards a smooth evolution from Rh$^{3+}$ at doping levels below $x=0.05$ towards Rh$^{4+}$ at doping levels above $x=0.70$. The Rh valence gradually increases towards the Rh$^{4+}$ value of the end compound Sr$_2$RhO$_4$. We note that the contraction of lattice parameters with doping in \Rhdoped, namely, chemical pressure~\cite{Qi2012a} cannot explain the observed leading edge shifts. For example, such {\it K}-edge shifts are not seen in SrRuO$_3$ under strong lattice compression achieved with external pressure.~\cite{Zhernenkov2013} Ab-initio simulations using the FDMNES code also show negligible leading edge shift for a hypothetical Sr$_2$RhO$_4$ structure with compressed lattice parameters.~\cite{Veiga2014} While {\it K}-edge shifts may be influenced by structural relaxation effects ~\cite{Vries2002, Herrero2010} we have no evidence that indicates these effects are at play in our data.

\begin{figure}
 \includegraphics[scale=0.28,angle= 0]{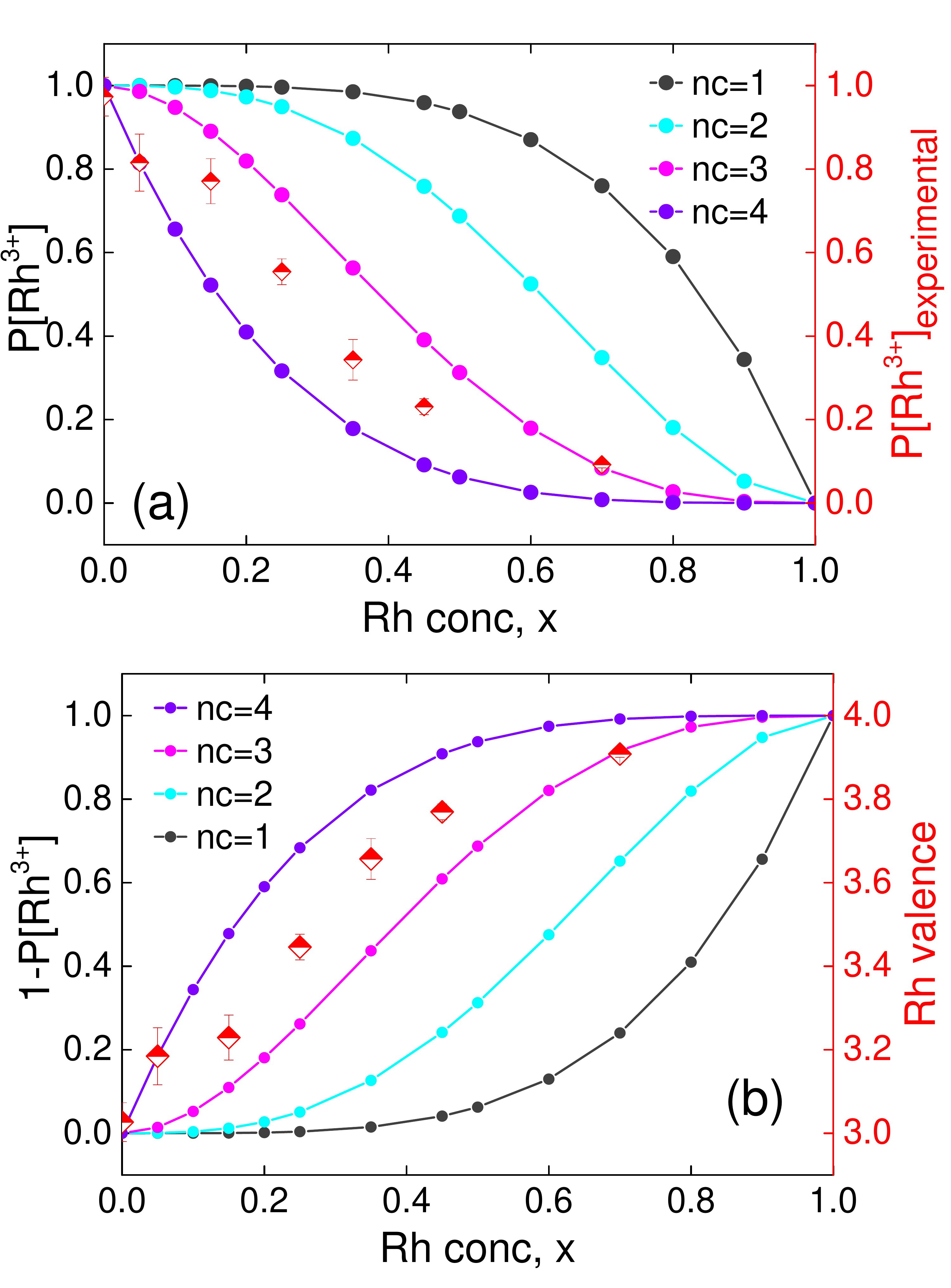}
 \caption{ (a) Calculated probability P[Rh$^{3+}$] for emergence of Rh$^{3+}$ state when the number of Ir neighbors is less than a critical number $n_c$ in the 1-4 range. On the right axis is the experimentally derived P[Rh$^{3+}$]. (b) shows P[Rh$^{4+}$] when the number of nearest neighbor Ir are below $n_c$. Right axis is the Rh valence from Fig.~\ref{RhKedge}(b). Experimental results imply that Rh is 3+ when the number of Ir neighbors exceeds a critical number between 3 and 4 (and Rh is 4+ when below this critical number).}
\label{Rh_probability}
\end{figure}

\begin{figure}
\includegraphics[scale=0.28,angle= 0]{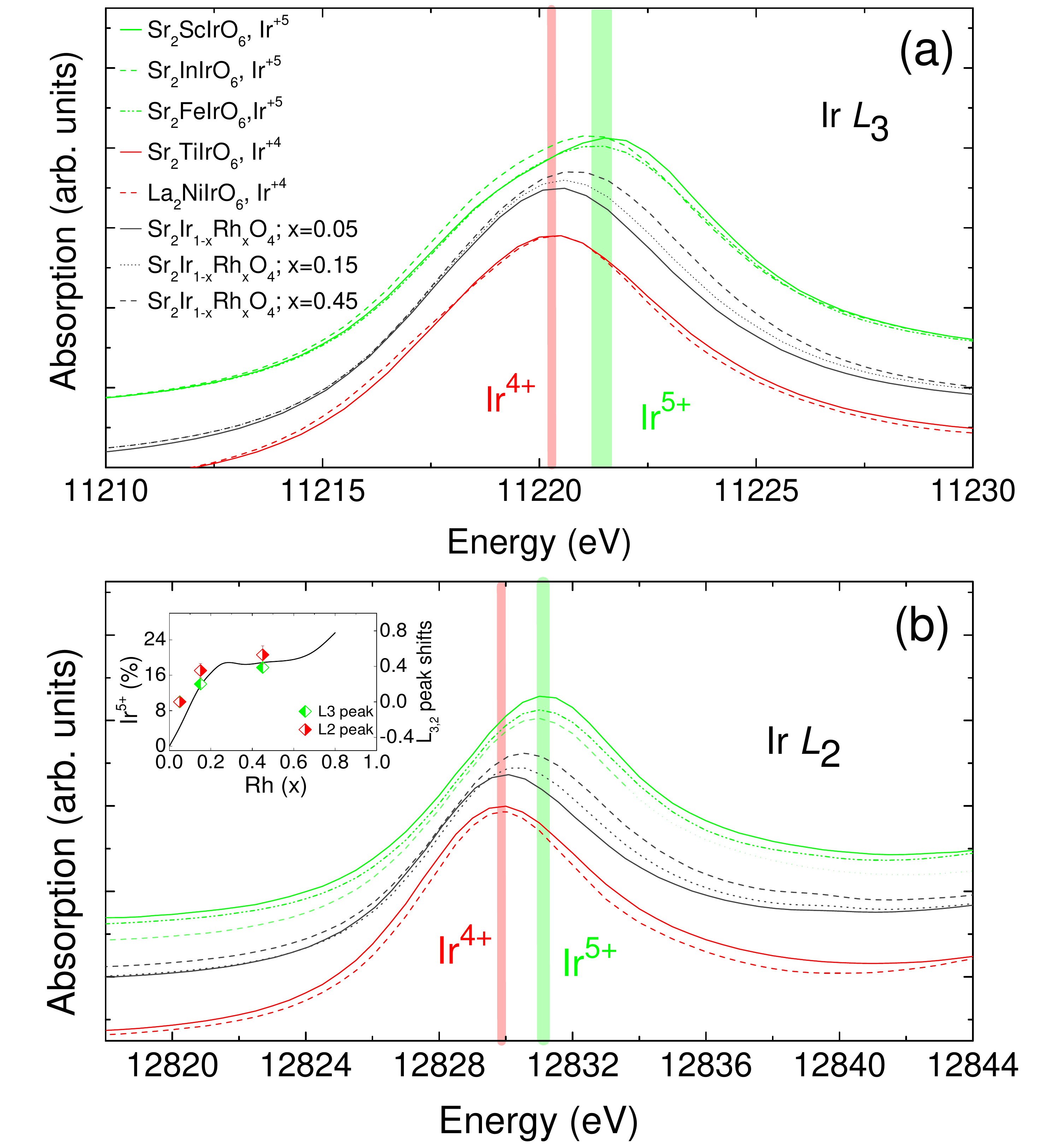}
\caption{(a) Ir $\rm{\it L}_3$ and (b) $\rm{\it L}_2$ XANES measurements for $x=0.05$, $x=0.15$ and $x=0.45$ samples together with Ir 4+ and 5+ standards. Data have been shifted in the vertical scale for clarity. Inset in (b) shows the $\% (\rm{Ir}^{5+})$ as a function of Rh, x on the left axis. The right axis shows the Ir $\rm{\it L}_{3,2}$ peak shifts for $x=0.05$, $x=0.15$ and $x=0.45$}
\label{Ir_edge} 
\end{figure}

In chemical compounds with more than one transition metal ion it is not uncommon to find charge disproportionation whereby charge transfer from heavier to the lighter ion takes place.~\cite{Streltsov2016, Meyers2014} If such charge partitioning takes place between Ir and Rh ions it will depend on the number of Ir/Rh nearest neighbors (n.n.) to a Rh ion. We work under the ansatz that the Rh valence $\nu_{Rh}(x)$,  will be 3+ when the number of Ir n.n. exceeds a critical value $n_c$.~\cite{Khomskii1979} We use a binomial distribution to calculate the probability, P[Rh$^{3+}$] for the number of Ir n.n. to exceed $n_c$ for different $x$ values. Fig.~\ref{Rh_probability} (a) shows the calculated probability P[Rh$^{3+}$] of Rh being in the 3+ state for different $n_c$ values as a function of $x$. The corresponding experimental value can be obtained from the experimental valence (here we used values derived from Rh {\it K}-edge data shown in Fig.~\ref{RhKedge} (c)), $\nu_{Rh}(x)$ = $3\times \rm{P[Rh^{3+}]} + 4\times \rm{(1-P[Rh^{3+}])}$. Fig.~\ref{Rh_probability} (b) shows the calculated probability of Rh being in the 4+ state, P[Rh$^{4+}$]=1-P[Rh$^{3+}$] when the number of Ir n.n. is less than $n_c$, displayed as a function of $x$. On the right axis we display the experimental Rh valence derived from Rh {\it K}-edge data and shown in Fig.~\ref{RhKedge}(b). We see that the experimental and calculated values are in reasonably good agreement when the critical number of Ir neighbors is between 3 and 4 for each Rh site. 

We have assumed that charge disproportionation takes place between Rh and Ir ions but the possibility that oxygen ions participate in charge compensation ought to be considered. A recent O $K$-edge XANES study using \Rhdoped\ samples grown under the same conditions as used in our study, however, has shown that there is no measurable charge transfer involving O 2p states.~\cite{Sohn2016}
Fig.~\ref{Ir_edge} shows normalized XANES at the Ir $\rm{\it L}_{3,2}$ edges for three representative \Rhdoped\ samples as well as tetravalent and pentavalent Ir standards.~\cite{Laguna-Marco2015} If charge transfer between Rh 4d and Ir 5d states takes place, we expect a concomitant evolution of Ir valence accompanying changes in Rh valence. In the presence of Rh$^{3+}$, the fraction of Ir ions in a 5+ valence state needed to achieve charge neutrality is simply given by $x\times\rm{P[Rh^{3+}]}$/$(1-x)$. This function, calculated using experimental P[Rh$^{3+}$] values in Fig.~\ref{Rh_probability}(a), is shown in the inset in Fig.~\ref{Ir_edge} (solid line, left axis). The right axis of the inset shows experimental peak shifts in Ir L$_{3,2}$ edge spectra for the three representative samples (shifts are relative to 4+ standards;  total peak shift per Ir unit charge is $1.04 \pm 0.17$ eV). The fraction of Ir 5+ ions remains below about 25\% for all levels of doping. At small Rh concentrations only a small fraction of Ir atoms compensate the charge and hence XANES measurements, which average over all Ir sites, are rather insensitive with the apparent Ir valence remaining close to $4+$ for $x\leq0.1$ (no measurable peak shift relative to $4+$ reference). As Rh doping increases, the fraction of Ir$^{5+}$ ions increases and a clear peak shift of about 0.4 eV towards the position of the Ir$^{5+}$ is seen for the $45\%$ Rh doped sample. The data in Fig.~ \ref{Ir_edge} provides strong evidence in favor of charge compensation at Ir sites. The fraction of Ir$^{5+}$ ions increases in a non-linear fashion with Rh doping. We note that the Ir $ L_{3,2}$ branching ratio is not expected to show significant changes with doping. The experimental branching ratio changes by only $10 - 12\%$ in going from Ir$^{4+}$ to Ir$^{5+}$.~\cite{Laguna-Marco2015} Since only less than 25\% of Ir ions are in a 5+ valence state at any $x$ value we expect the branching ratio to be rather insensitive to Rh content in agreement with previous reports~\cite{Chikara2015}.
 
\begin{figure}
 \includegraphics[scale=0.35,angle= 0]{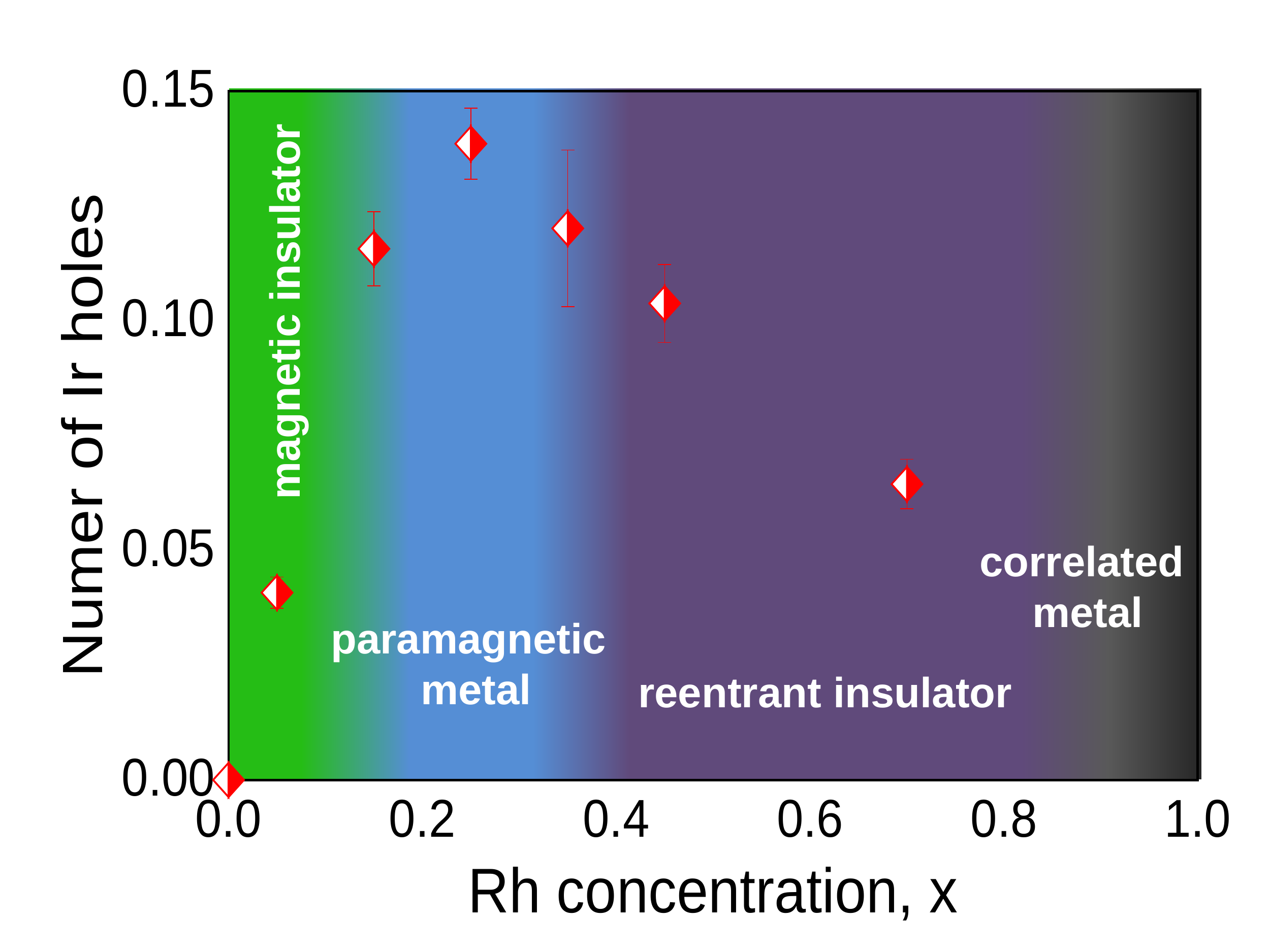}
 \caption{Number of 5d Ir holes calculated empirically using Rh valence values derived from Rh K-edge data in Fig.~\ref{RhKedge} are shown superimposed on the low temperature phase diagram adapted from Qi. et. al.~\cite{Qi2012a}}
\label{phase_diagram}
\end{figure}

The charge transfer from Ir 5d to Rh 4d orbitals via intervening oxygen atoms is consistent with density functional calculations showing that Rh occupied states in the vicinity of the Fermi level (lower Hubbard band) lie lower in energy than Ir 5d occupied states at low Rh doping levels.~\cite{Chikara2015} With increasing Rh doping, however, these 4d states move up in energy at a faster pace than 5d states as a result of increased  4d bandwidth and related reduction in on-site Coulomb interactions. This makes charge transfer between Ir and Rh d states less favorable at higher doping levels with a concomitant evolution of Ir/Rh valence towards the 4+ valence state.~\cite{Chikara2015} Charge transfer from heavier to lighter d-elements of the same column in the periodic table is also in agreement with the observations in CaCu$_3$B$_4$O$_{12}$ (B=Co, Rh, Ir) where Cu valence changes from predominantly 3+ to 2+ (Cu is reduced) as one moves from 3d to 4d to 5d elements at the B site.~\cite{Meyers2014}

The average number of doped 5d (Ir) holes per Ir site as a function of Rh content can be calculated as $x\times (4-\nu_{Rh}(x))$. This is shown in Fig.~\ref{phase_diagram} superimposed on the low temperature phase diagram reported by Qi et. al.~\cite{Qi2012a} The number of 5d Ir holes initially increases for $x\leq 0.25$ and then decreases on further doping. At low Rh concentrations Rh assumes a predominant 3+ state which dopes holes into the Ir 5d band and explains the rapid decrease in the resistivity of the system leading to a metallic state. No superconducting behavior has been observed up to 15 at.\% Rh doping where the system is purely hole doped. With increasing $x$, the number of doped holes decreases systematically as the average Rh valence moves towards 4+. A reentrant insulating phase emerges at about $x\sim 0.35$ when the number of holes drops below about 0.12. Interestingly, this hole content threshold appears to be the same threshold needed to drive the metallic phase at $x\sim 0.16$. We note that the electrical resistivity of the insulating phase at low $x$ and that in the reentrant insulating phase shows similar temperature dependence~\cite{Qi2012a} pointing to a similar mechanism for localization of carriers in both phases. The decrease in the number of charge carriers is the most likely explanation for the reentrant insulating phase although doping induced disorder is also likely to be at play.

In summary, combining Rh {\it K-, L-} and Ir {\it L-}edge XANES data we have shown that the effective Rh valence evolves from 3+ towards 4+ with increasing Rh doping. Charge compensation takes place at the Ir sites. Only a small fraction of Ir ions (less than $25\%$) assume a 5+ oxidation state across the doping-dependent Rh/Ir charge partitioning. The number of holes doped into the Ir 5d J$_{\rm eff}$=1/2 band, therefore, shows anomalous behavior increasing at low $x$ and driving an insulator-metal transition only to decrease at higher $x$ where a reentrant insulating phase is observed. This unexpected electronic doping should be taken into account when searching for novel electronic phases in iridates doped with 4d (or 3d) elements where charge disproportionation is expected.\\

\subsection*{Acknowledgement}
The authors thank Dr. Sergei Streltsov for insightful discussions. The authors would like to thank Dr. Larissa Veiga for the FDMNES calculations. The work at the Advanced Photon Source, Argonne National Laboratory, is supported by the U.S. Department of Energy, Office of Science under Grant No. DEAC02-06CH11357. Work at Los Alamos National Laboratory is supported by the Laboratory-Directed Research and Development program (LDRD). The NHMFL Pulsed-Field Facility is funded by the U.S. National Science Foundation through cooperative Grant No. DMR-1157490, the State of Florida, and the U.S. Department of Energy. Work at Brookhaven National Laboratory is supported by the US Department of Energy, Office of Basic Sciences, Early Career Award Program under Award Number 1047478. GC acknowledges support of U.S. National Science Foundation grants DMR-1265162 and DMR-1600057. The work of D.Kh. was supported by the German Excellence Initiative and by the SFB 1238. 

\bibliography{Rh_valence_paper}

\end{document}